# Doping Human Serum Albumin with Retinoate Markedly Enhances Electron Transport Across the Protein

Nadav Amdursky[a,b], Israel Pecht[c], Mordechai Sheves[b,*], and David Cahen[a,*]

*Departments of [a]Materials and Interfaces, [b]Organic Chemistry, and [c]Immunology, Weizmann Institute of Science, Rehovot 76100, Israel*

**Abstract**

Electrons can migrate via proteins over distances that are considered long for non-conjugated systems. Proteins' nano-scale dimensions and the enormous flexibility of their structures and chemistry makes them fascinating subjects for investigating the mechanism of their electron transport (ETp) capacity. One particular attractive research direction is that of tuning their ETp efficiency by doping them with external small molecules. Here we report that solid-state ETp across human serum albumin (HSA) increases by more than two orders of magnitude upon retinoate (RA) binding to HSA. RA was chosen because optical spectroscopy has provided evidence for the non-covalent binding of at least three RA molecules to HSA and indications for their relative structural positions. The temperature dependence of ETp shows that both the activation energy and the distance-decay constant decrease with increasing RA binding to HSA. Furthermore, the observed transition from temperature-activated ETp above 190K to temperature-independent ETp below this temperature suggests a change in the ETp mechanism with temperature.

Bridging the worlds of biology and solid-state electronics presents a fascinating challenge.[1] One approach to achieve this is exploring the use of biomolecules as electron transport (ETp) materials.[1b] Indeed, diverse biological macromolecules, from proteins,[2] peptides,[3] to DNA,[4] and peptide-nucleic acids[5] have been studied as such, using scanning probe (conductive-probe atomic force or scanning-tunneling) microscopy or measurements via molecules, sandwiched between macroscopic electrodes.[1b, 2a, 2c-e, 6] Most studied proteins are those with biological electron transfer (ET) function, such as azurin,[2c, 2e, 7] cytochrome C[6, 8] or plastocyanin.[9] We have recently shown, using macroscopic contacts, that bacteriorhodopsin (bR), a protein that does not have a natural ET function, but contains the covalently bound cofactor retinal, has a room temperature ETp efficiency similar to that of ET proteins.[2c, 10] However, proteins devoid of any cofactor, such as apo-azurin, serum albumin, as well as washed apo-bR, exhibit an order of magnitude lower ETp efficiency than, e.g., the ET protein azurin.[2c, 7b]

Here we show that ETp via human serum albumin (HSA), a protein that lacks any cofactor, increases by over 2 orders of magnitude upon binding non-covalently *(doping with)* 3 equivalents of deprotonated retinoic acid, i.e., retinoate, (RA). RA is a close derivative of the bR cofactor (retinal), and has a biological role in the growth and development of the embryo.[11] We chose to study HSA, because of its extraordinary binding capacity of small, mainly hydrophobic, molecules.[12] We used the all-trans isomer of RA, due to its relatively high affinity to HSA, with an aneraged binding constant of $3.3 \cdot 10^5$ $M^{-1}$, and as it has no known effect on the proteins' conformation.[13] HSA has been suggested to bind up to three equivalents of RA,[13] but as no three-dimensional structure of an HSA-RA complex is available, the locations of the RA binding sites within HSA are not agreed on. Kossi et al.[13b] suggested that RA binding sites are those, known for long-chain fatty acids.[14] Maiti et al.[13d] proposed, based on molecular dynamics docking simulations, one hydrophobic site between subdomains IIA and IIIA, while Belatik et al.[13e] suggested the site of subdomain IB (one of the binding sites for long-chain fatty acids) to be the main RA binding site (cf. Fig. S1 in the supporting information).

The addition of RA (stock solution in ethanol) to HSA (aqueous PBS buffer, pH 8) leads to its binding, which can be monitored by UV-Vis absorption spectroscopy.[13] As illustrated in Fig. 1, a gradual increase in absorption is observed ($\lambda_{max}$=346 nm, ascribed to retinoate[13c]) as the ratio of ligand to protein (L/P), i.e., RA/HSA, was increased, suggesting RA-HSA complex formation. As also seen in the figure, at L/P = 4, a second band (at ~405 nm), red-shifted from the main band (346 nm) appears. This second band is ascribed to retinoic acid.[13c] The increase in the 346 nm band intensity at L/P = 4 suggests that a fourth RA might bind to HSA, but due to the appearance of the second absorption band at ~405 nm we can conclude that the pKa of this binding site is somewhat increased in comparison to the other binding sites. However, and more important to the scope of this communication (as will be shown later), the fourth equivalent hardly affects the ETp characteristics. Further evidence for retinoate being the main bound species to HSA is derived from the absorption spectra of RA-HSA complexes with different L/P ratios at different pH values (Fig. S2, and supplemental text), which show lower absorption at low pH's (pH = 6-7) than at high pH's (pH = 8-9).

Monolayers of HSA alone and of its complexes with RA at different L/P ratios were prepared on a Si surface for ETp measurements. The used protocol was the same as that previously employed for studies of bovine serum albumin (BSA).[2c] The similarity of surface coverage by HSA at different L/P ratios was verified optically by ellipsometry, yielding an optical thickness of 17-20 Å. In addition, atomic force microscopy indicated similar surface morphology of the different surfaces (Fig. 2a), comparable to that found for surfaces of BSA monolayers (Fig. S3).[2c]

While surface coverage and morphology with different L/P were similar, the ETp efficiency *at room-temperature* (Fig. 2b) showed up to a nearly 2 orders of magnitude increase with increasing equivalents of bound RA. The main increase (an order of magnitude) was observed upon binding the third equivalent, while addition of the fourth one only gave a small further increase.

To investigate the ETp mechanism we studied the change in the current density as a function of temperature (Fig. 3a). As seen in the figure, all curves exhibit a temperature-independent regime at low temperatures (<190K), and a thermally activated regime at higher temperatures (≥190K). The magnitude of the current densities and their change as a function of temperature are quite similar for HSA and BSA (Fig. S4). The current den-sities across the proteins can be presented by different expressions, for the temperature-independent regime, $J_{TI}$, and for the thermally activated regime, $J_{TA}$:[10, 15]

$$J_{TI} \propto \exp(-\beta l) \qquad J_{TA} \propto \exp(-E_a / k_B T) \qquad (1)$$

where $\beta$ is the distance-decay constant of ETp, $l$ is the geometrical separation of the electrodes across which ETp takes place, $E_a$ is the activation energy for the thermally activated ETp, $k_B$ is the Boltzmann constant and $T$ is the absolute temperature. The dominant process at a certain temperature determines the measured current density. We will now discuss the effect of binding RA to HSA on ETp in these two regimes.

At low temperatures, where $J_{TI}$ dominates, we find a ~330 fold increase in current densities for the HSA complex with three RA equivalents compared to those of HSA alone. The left proportionality of Eq. 1 implies that the change in the current density can be explained by either a change in $\beta$ (assuming that the distance between the electrodes remains the same) or by the pre-exponential factor (not shown in the proportionality), which refers to the coupling matrix of the ETp process via the electrode-protein-electrode system. While we cannot estimate the change in the coupling matrix, the similarity in the junction configurations and the constant separation distance between the electrodes favors $\beta$ as the main parameter that changes. With this assumption the *increase* in temperature-independent current densities can be expressed as a result of a *decrease* in $\beta$ values:

$$\frac{J_{TI,0}}{J_{TI,3}} = \frac{H \exp(-\beta_0 l)}{H \exp(-\beta_3 l)} = 3 \times 10^{-3} \xrightarrow{l \approx 37 \text{Å}} \beta_3 [\text{Å}^{-1}] = \beta_0 - 0.16. \qquad (2)$$

Here $H$ is the coupling matrix, $J_{TI,n}$ and $\beta_n$ are the current density in the temperature-independent regime and the distance-decay constant at L/P = $n$, respectively. For the calculation we took the thickness of the protein monolayer (perpendicular to the surface), i.e., the maximal distance across which ETp occurs, to be ~37 Å, as determined by the 3-dimensional structure of the protein (PDB ID: 1E7I, Fig. S5), and also illustrated by the z-axis of the AFM image (Fig. 2a). In a similar way we can calculate that $\beta_2 = \beta_0 - 0.07$ and $\beta_1 = \beta_0 - 0.05$. We assume that in this temperature regime ETp occurs by resonant tunneling, provided by the bound RA, and even more so that of several proximally bound

RAs, within HSA. The presence of RA lowers the energy levels, involved in the ETp process, similar to what is observed in redox proteins,[16] explaining the decrease in $\beta$ values.

Binding RA to HSA also affects the ETp behavior in the high-temperature regime, where thermally activated ETp is observed (current density described by $J_{TA}$, as in eq. 1), with clear differences in the currents' temperature dependence at different L/P ratios. Plots of $\ln(J_{TA})$ vs. $T^{-1}$ are linear (Arrhenius plots; Fig. 3b) and show that the activation energy, $E_a$, decreases as L/P increases. Here, too, it is reasonable to assume that the introduction of several RAs lowers the energy barriers for the non-adiabatic ETp process, corresponding to shallower diabatic curves and lower reorganization energies, which, consequently cause a decrease in the observed activation energy.

The large increase in current density through the protein monolayer as a function of RA binding is remarkable, especially because the doped protein has no known electron or charge transfer role in nature. This emphasizes the crucial role of any bound cofactor in the ETp via proteins. It also suggests that the structure of a protein can serve as framework for efficient ETp, upon binding an appropriate cofactor, irrespective of the protein's natural function. In this context, we note that the current density (at 0.05 V) through the doped HSA monolayer (at L/P = 3) is even slightly higher, especially in the low-temperature regime, than through bR, a natural retinal-containing protein,[10] or through a natural ET protein, azurin.[7b]

In conclusion, we discovered that the ETp via a solid-state monolayer of HSA increases dramatically as a result of HSA doping with up to 3 equivalents of RA. The increase in ETp with increasing L/P can be interpreted as a change in ETp parameters. This change is expressed as a decrease both in the distance-decay constant of the temperature-independent ETp at low temperatures, and in the activation energy of the temperature dependent ETp at higher temperatures. The drastic increase in currents by RA-doping of HSA confirms and significantly extends our earlier *tenet* that proteins can behave as molecular wires, upon cofactor binding. The ability to markedly enhance the electrical conductivity of a protein monolayer over a significant range may be of interest for using proteins as ETp mediators in future bio-electronics devices.

**Figures**

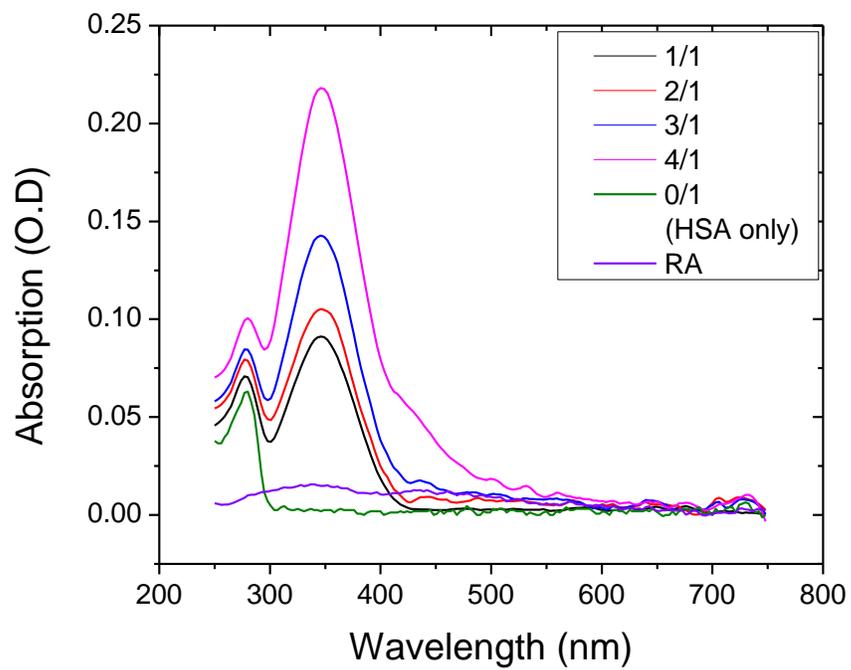

**Figure 1**. Optical UV-Vis absorption spectra of HSA and RA at different L/P ratios at pH=8.

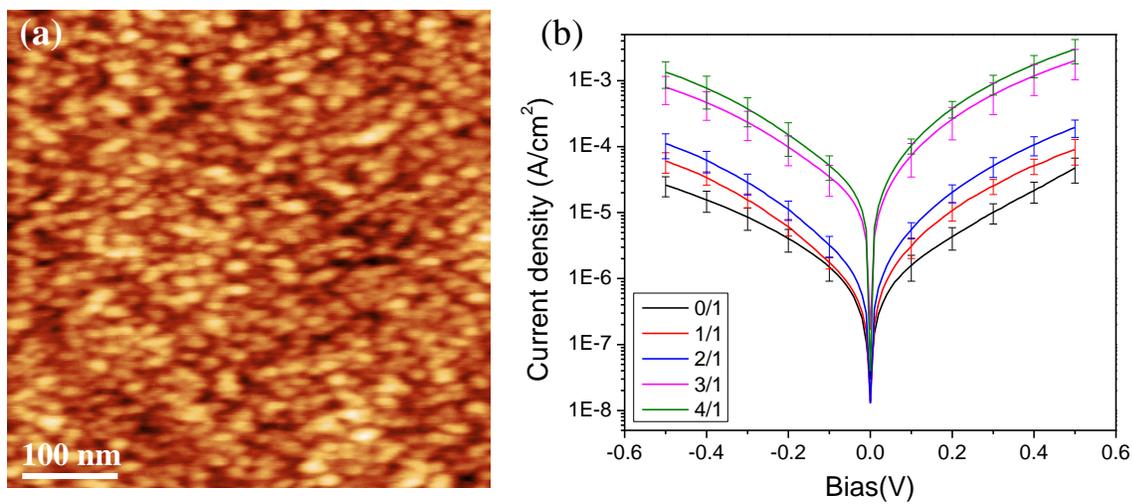

**Figure 2**. Solid state monolayer on Si substrate. (a) AFM image of an HSA monolayer surface. Z-scale is 4.5 nm; the rms roughness is centered at 1.7 nm. (b) Current density-voltage curves of HSA-RA complexes with L/P = 0 - 4, at room temperature.

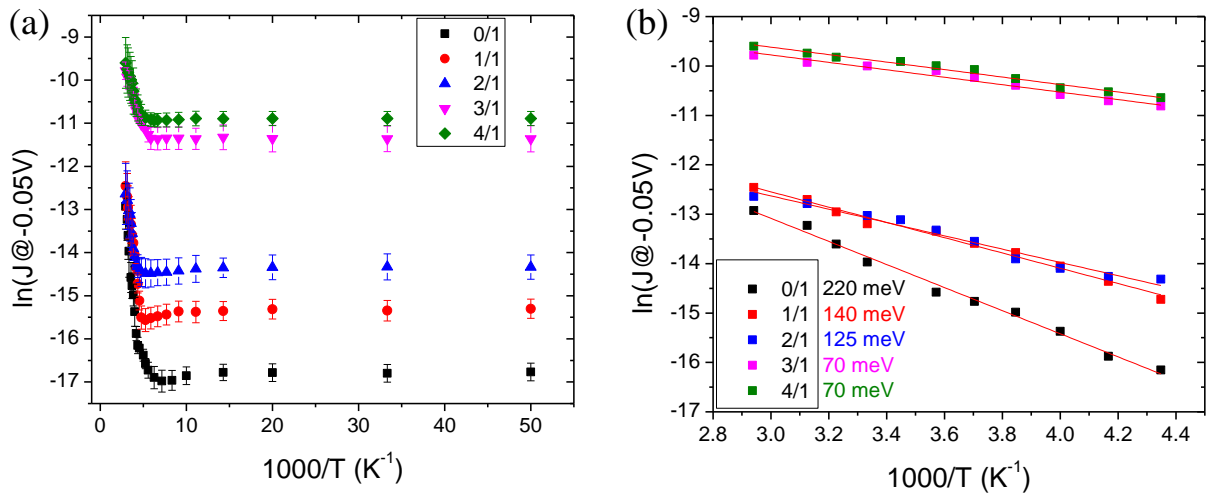

**Figure 3**. ETp temperature dependence. (a) current density at -50 mV as a function of temperature over the range of 20-340 K. (b) zoom of (a) in the thermally-activated regime (230-340K), showing also the calculated activation energies for each L/P ratio.

**Supplementary Material**

for

**Doping Human Serum Albumin with Retinoate Markedly Enhances Electron Transport Across the Protein**

*Nadav Amdursky[a,b], Israel Pecht[c], Mordechai Sheves[b,*], and David Cahen[a,*]*

*Departments of [a]Materials and Interfaces, [b]Organic Chemistry, and [c]Immunology, Weizmann Institute of Science, Rehovot 76100, Israel*

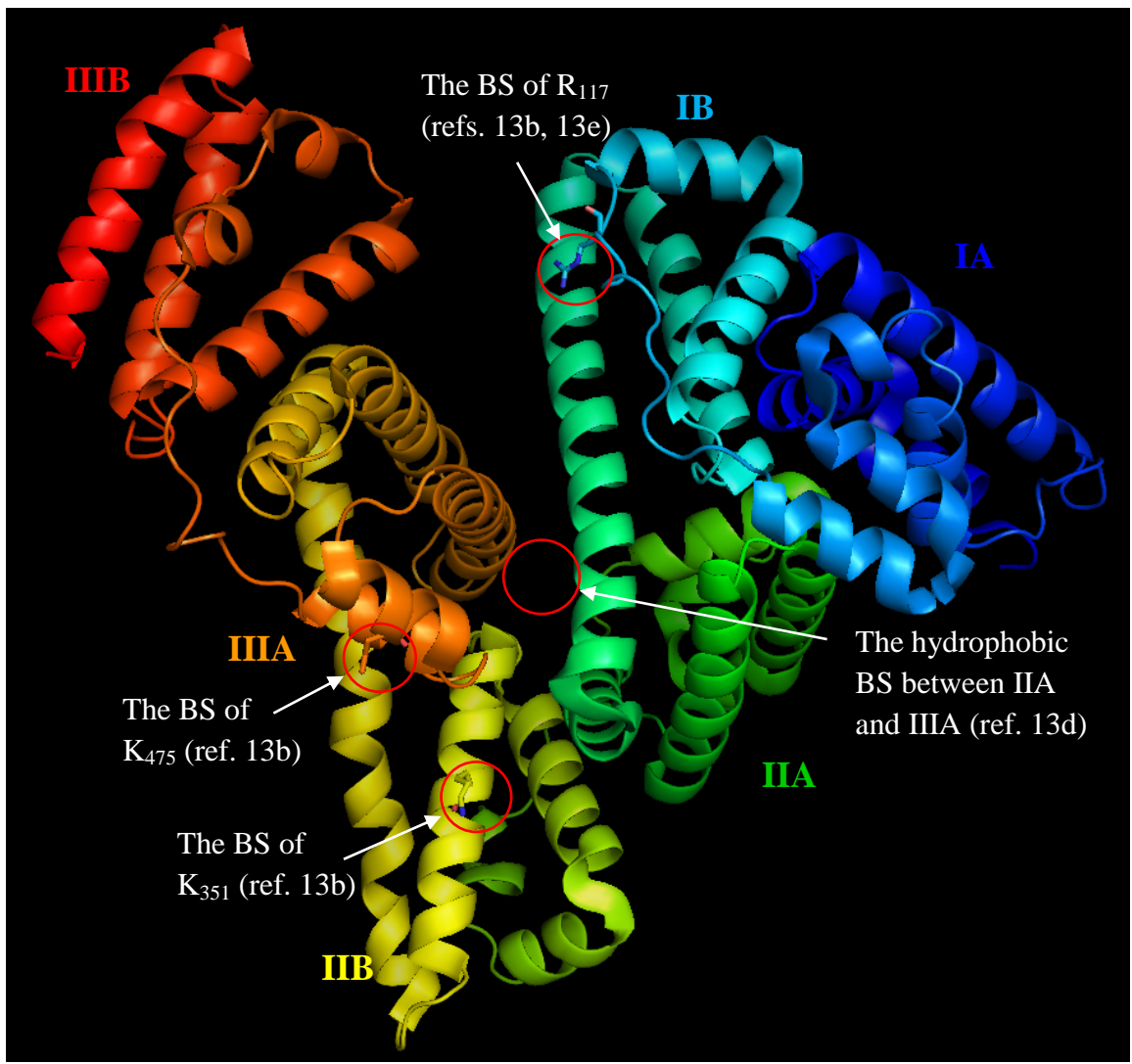

**Figure S1**. The structure of HSA (PDB: 1E7I) with its sub-domains, along with the different suggested RA binding sites (references are given in the picture).

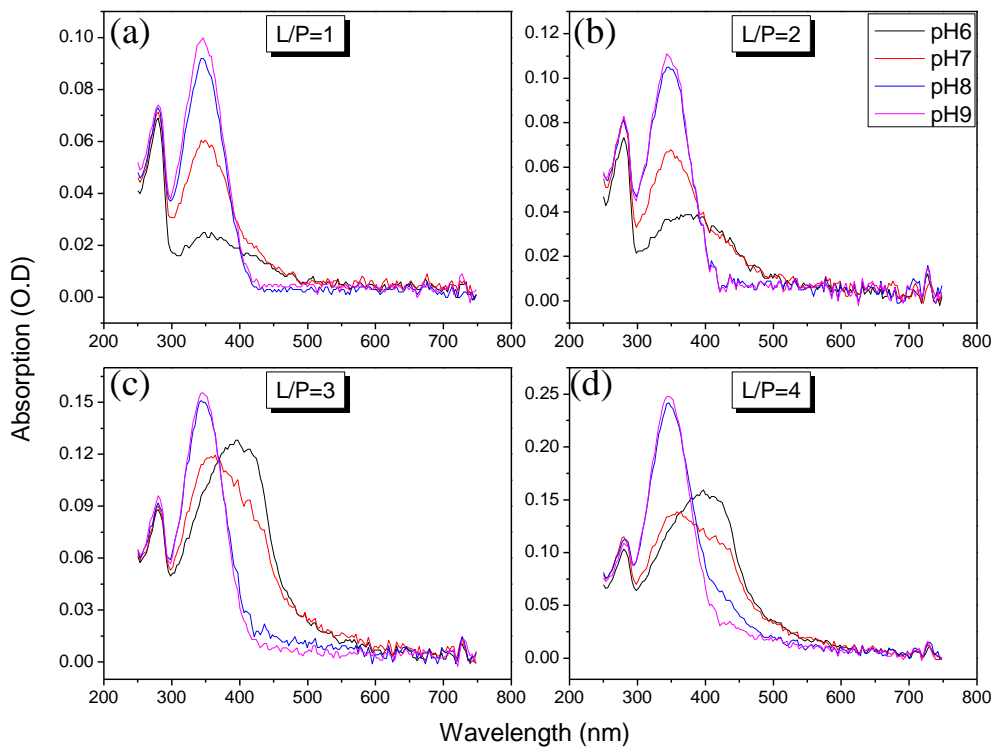

**Figure S2**. Optical UV-Vis absorption spectra of HSA and RA at Different pH's (pH 6-9) at L/P = 1, 2, 3 and 4, (a) − (d) respectively.

**Binding of retinoate to HSA**

Figure S2 shows the lower absorption of RA (or retinoic acid) in the presence of HSA at low pH's (pH = 6-7) than at high pH's (pH = 8-9). Another indication for inefficient binding at pH = 6-7 is that already upon addition of the first RA equivalent a turbid yellowish solution results, whereas at pH = 8-9 the solution remains clear up to L/P = 3. The observation that retinoate binds to HSA, supports the binding sites location proposed previously,[1] since these binding sites include positively charged amino acids, which will enable retinoate binding via electrostatic interaction. As the optical absorption spectra of RA binding to HSA do not allow for distinguishing between the different RA binding sites, we cannot exclude the possibility of having a mixture of proteins with L/P ratio, different from that of the solution.

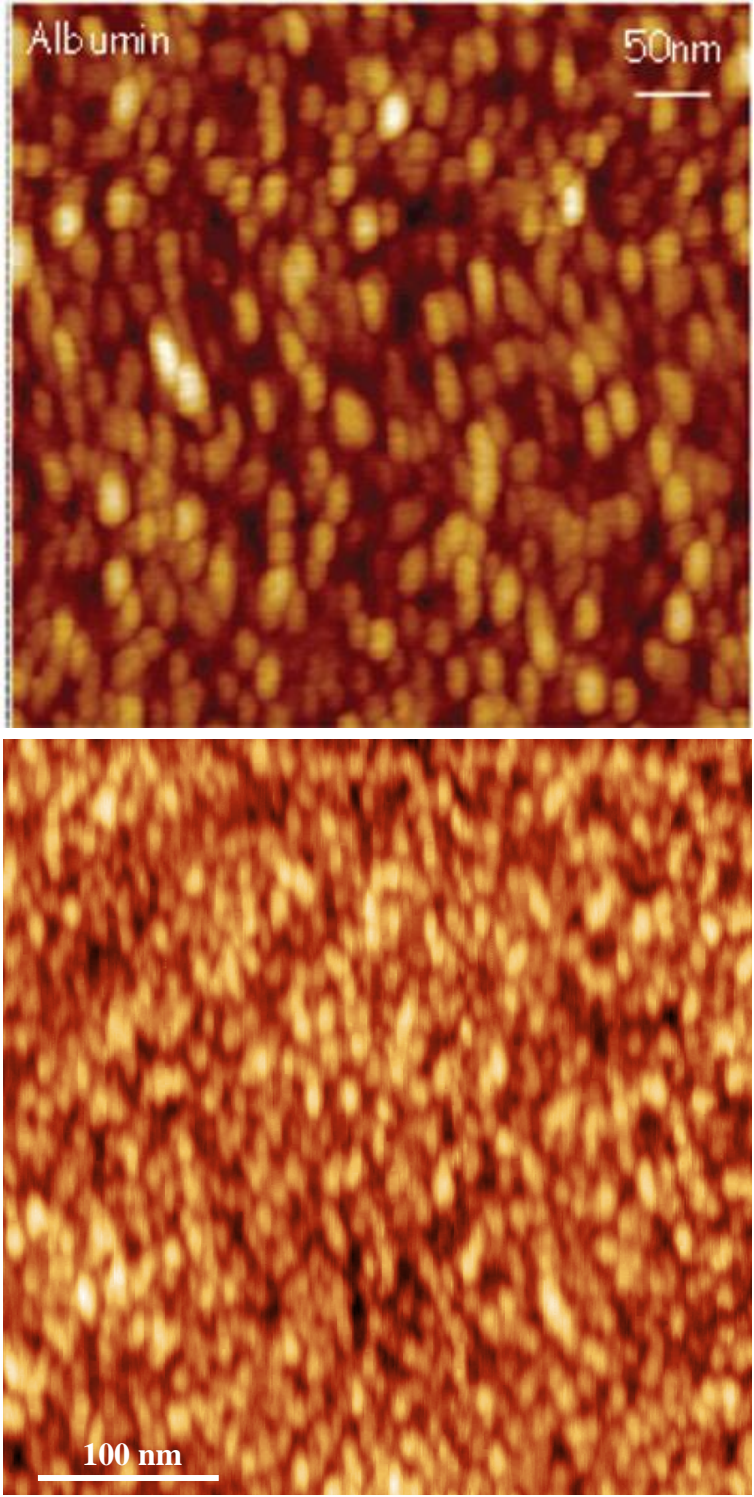

**Figure S3**. Comparison of the surface morphology, as measured by AFM, between BSA (top image, taken from ref. 2c) and HSA (bottom image).

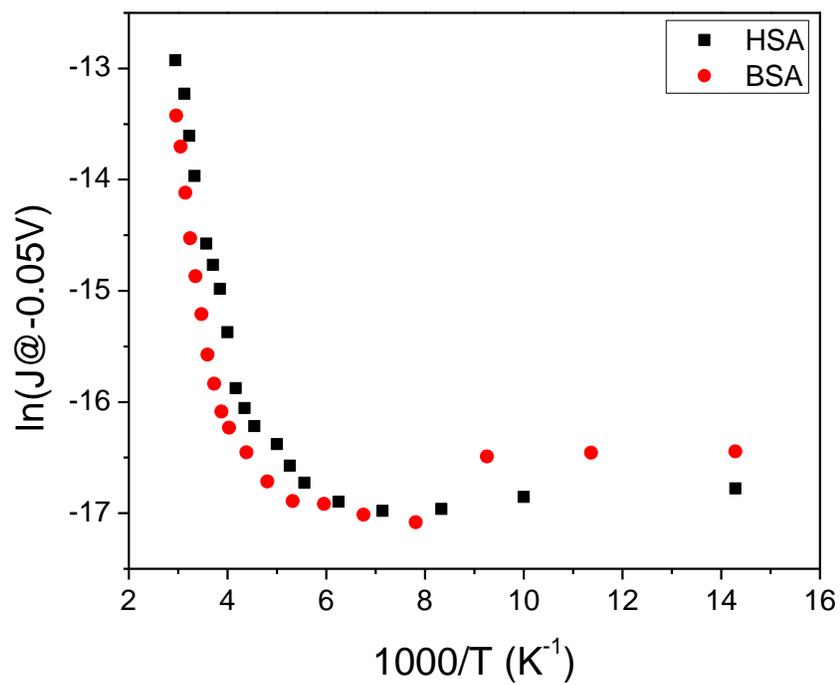

**Figure S4**. Comparison of the current density (at -0.05V) as a function of temperature via BSA (taken from ref. 7b) and via HSA monolayers.

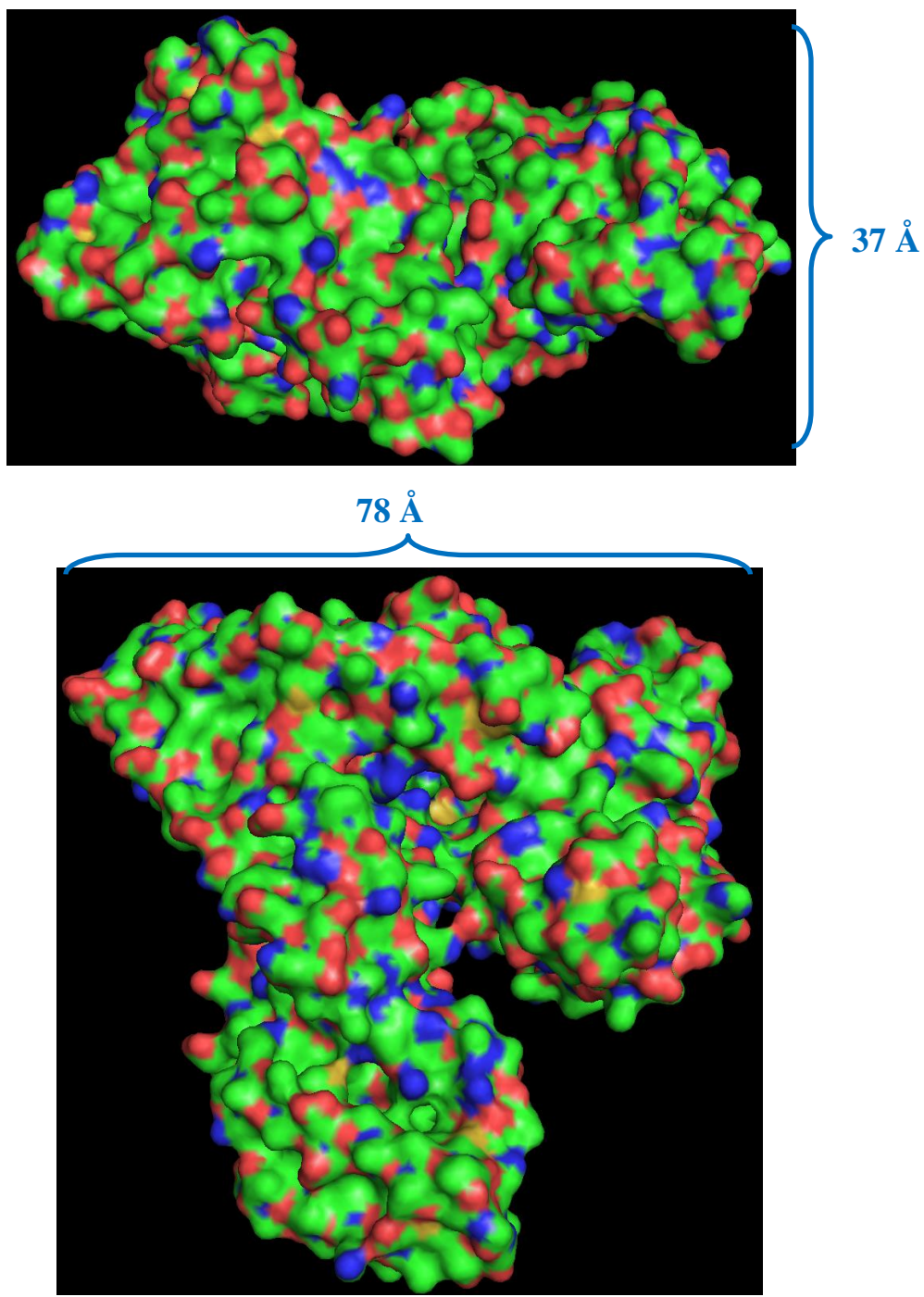

**Figure S5**. Top image: the short axis of HSA, which assumed to be the thickness of the monolayer (the protein is perpendicular to the surface). Bottom image: the long axis of HSA.